\documentclass[aps,prl,twocolumn,preprintnumbers,amsmath,amssymb,superscriptaddress]{revtex4}

\usepackage{graphicx}

\usepackage{pifont} 
\usepackage{wasysym} 

\begin{document}

\title{Loss and decoherence due to stray infrared light in superconducting quantum circuits}

\author{R. Barends}
\author{J. Wenner}
\author{M. Lenander}
\author{Y. Chen}
\author{R. C. Bialczak}
\author{J. Kelly}
\author{E. Lucero}
\author{P. O'Malley}
\author{M. Mariantoni}
\author{D. Sank}
\author{H. Wang}
\author{T. C. White}
\author{Y. Yin}
\author{J. Zhao}
\author{A. N. Cleland}
\author{John M. Martinis}
\affiliation{Department of Physics, University of California, Santa
Barbara, CA 93106, USA}
\author{J. J. A. Baselmans}
\affiliation{SRON Netherlands Institute for Space Research,
Sorbonnelaan 2, 3584 CA Utrecht, The Netherlands}

\date{\today}

\begin{abstract}
We find that stray infrared light from the 4~K stage in a cryostat
can cause significant loss in superconducting resonators and qubits.
For devices shielded in only a metal box, we measured resonators
with quality factors $Q = 10^5$ and qubits with energy relaxation
times $T_1 = 120$~ns, consistent with a stray light-induced
quasiparticle density of 170-230 $\mu$m$^{-3}$. By adding a second
black shield at the sample temperature, we found about an order of
magnitude improvement in performance and no sensitivity to the 4~K
radiation. We also tested various shielding methods, implying a
lower limit of $Q = 10^8$ due to stray light in the light-tight
configuration.
\end{abstract}

\maketitle

Quantum information processing in superconducting circuits is
performed at very low temperatures, so energy loss due to
quasiparticles is expected to vanish because their density
diminishes exponentially with decreasing temperature. As energy
relaxation times saturate for superconducting quantum circuits and
planar resonators, reaching values on the order of 1-10 $\mu$s
\cite{houck,bertet,wang2009,barends2010}, recent experiments have
suggested that this may be due to excess non-equilibrium
quasiparticles; measurements on phase qubit coherence
\cite{martinis2009,lenander}, tunneling in charge qubits
\cite{shaw2008}, resonator quality factors
\cite{wang2009,barends2010} and quasiparticle recombination times
\cite{barends2009,visser2011} are compatible with an excess
quasiparticle density on the order of 10-100 $\mu$m$^{-3}$, possibly
arising from stray light, cosmic rays, background radioactivity, or
the slow heat release of defects.

\begin{figure}[b]
    \centering
    \includegraphics[width=1\linewidth]{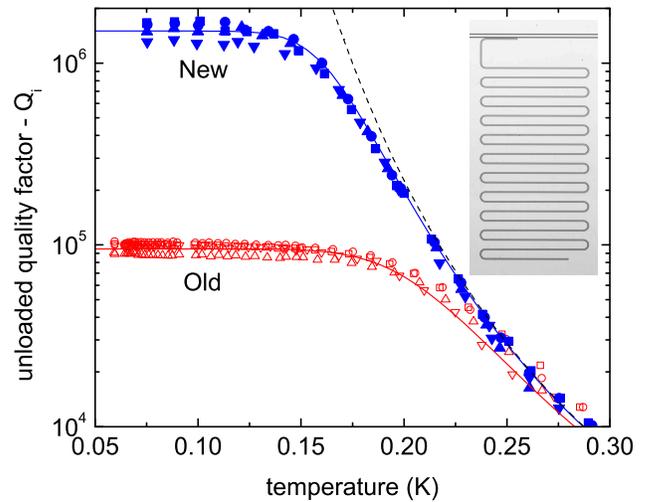}
    \caption{(Color online) The quality factors of four halfwave coplanar waveguide Al resonators versus sample stage temperature,
    measured in a setup without effort to shield stray infrared light (open symbols) and with an improved light-tight sample stage (closed symbols).
    Resonance frequencies lie between 3.8 and 4.5 GHz.
    Eq. \ref{eq:lossvsnqp} is plotted for an exponentially decreasing quasiparticle density (dashed line),
    excess quasiparticle density of 230 $\mu$m$^{-3}$ (bottom solid line) and 10 $\mu$m$^{-3}$ (top solid line).
    Kinetic inductance fraction $\alpha=0.28$ for these devices. Inset shows a halfwave resonator capacitively coupled to a feedline.}
    \label{figure:fig1}
\end{figure}

In this Letter, we demonstrate that stray infrared light gives
significant loss in resonators and qubits, and is sometimes the
dominant limitation in our present experiments. We also show
quantitatively how a combination of infrared shielding techniques
removes the influence of stray infrared light, and that the required
shielding is beyond what is generally used. The effectiveness of the
various techniques is investigated by methodically changing and
testing them. With our new light-tight setup, the quality factors of
Al superconducting resonators improve dramatically by a factor of
20, as shown in Fig.\,\ref{figure:fig1}. We also show that shielding
improves phase qubit coherence.

Loss in a superconducting resonator with frequency $f$ is controlled
by the quasiparticle density $n_{qp}$ \cite{mattis} (for $kT\ll hf$)
\begin{align}
\label{eq:lossvsnqp}
\frac{1}{Q} = \frac{\alpha}{\pi} \sqrt{\frac{2\Delta}{hf}} \frac{n_{qp}}{D(E_F)\Delta}
\end{align}
with $\Delta$ the energy gap, $D(E_F)$ the two-spin density of
states, and $\alpha$ the kinetic inductance fraction, which depends
on geometry. Importantly, excess quasiparticles can limit quality
factors, in particular at the low temperatures at which resonators
and qubits are operated.

Measurements on the temperature dependence of resonator quality
factors indicate the presence of an additional loss term, as shown
in Fig. \ref{figure:fig1}. Here we plot quality factors of coplanar
waveguide (CPW) Al resonators. The open symbols are measured when
simply placing the sample in a sample box in a cryostat, with no
special measures against stray light. Above a temperature of 200 mK
the quality factors decrease exponentially, consistent with a
thermal quasiparticle density (dashed line, Eq. \ref{eq:lossvsnqp}).
At low temperatures a plateau value of $10^5$ is observed,
consistent with an excess quasiparticle density of 230 $\mu$m$^{-3}$
(red solid line). When using our newly designed sample stage,
quality factors of the \emph{same} resonators improve to $2\cdot
10^6$ (closed symbols). This shows that stray infrared light is
limiting the quality factors in the old design, and that our new
design is light-tight. We will discuss our light-tight setup, test
the effectiveness of its parts, and show that the influence of stray
light can be fully removed.

Stray infrared light enters the sample mount through the lid joint
and connectors, generating quasiparticles in the sample. It can be
shown that under strong loading \cite{rothwarf}: $n_{qp} \propto
\sqrt{ P/\Delta }$, where $P$ is the absorbed radiation power for
which $hf>2\Delta$. Al is particularly sensitive to stray light:
with a gap frequency of 88 GHz, 96 \% of the power of a 4.2 K
blackbody can be absorbed. Moreover, quasiparticle recombination is
slow in Al \cite{barends2009,visser2011}.

In order to quantify the influence of stray light, we use halfwave
CPW Al resonators with a film thickness of 52~nm, which are coupled
capacitively to a feedline (inset Fig. \ref{figure:fig1}). This
allows us to extract the unloaded quality factor $Q_i$ from the
feedline transmission. We measure quality factors at high power to
reduce the influence of two-level systems
\cite{wang2009,barends2010}. We use halfwave resonators, where the
central line is galvanically isolated, to rule out quasiparticle
outdiffusion and hot electrons. Measurements are done in an
adiabatic demagnetization refrigerator (ADR), with a base
temperature of 50 mK. The sample space is shielded by a cryogenic
magnetic shield, attached to the 4~K stage. The transmission is
measured using a vector network analyzer, a low noise cryogenic and
room temperature amplifier.

\begin{figure}[t]
    \centering
    \includegraphics[width=0.85\linewidth]{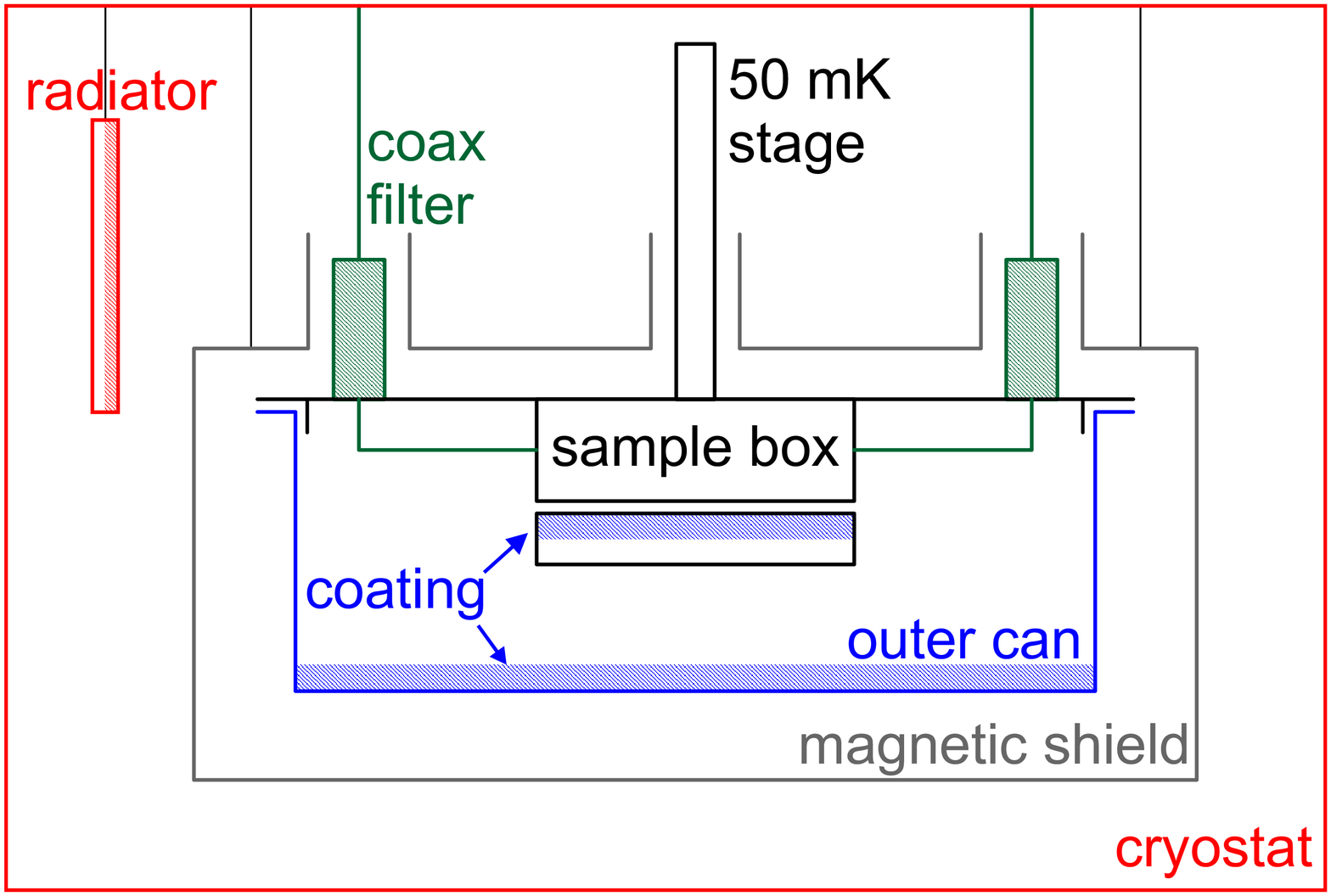}
    \caption{(Color online) Schematic representation of the light-tight sample stage:
    the sample box is mounted inside a larger box on the 50 mK stage, closed off by an outer can.
    The inner surfaces of the sample box lid and outer can are coated with a blackbody absorber (blue).
    Coaxial readout lines are filtered using 50 $\Omega$ matched metal powder filters (green).
    The entire sample stage lies within a magnetic shield (grey), attached to the cryostat's 4 K stage.
    The radiator is used for Fig. \ref{figure:fig4}.}
    \label{figure:fig2}
\end{figure}

Our light-tight sample stage uses a `box-in-a-box' design, following
Baselmans \emph{et al.} \cite{baselmansstray}. A schematic
representation, shown in Fig. \ref{figure:fig2}, gives a maximally
light-tight design. The sample box is placed in a larger box in
which the \emph{photon} temperature is equal or very close to the
desired \emph{electron} temperature. This is achieved by blocking
routes for stray light to enter the sample space as well as using
black coating on the inner surfaces. The black coating is a key
ingredient, and consists of a mixture of silica powder, fine carbon
powder and 1 mm SiC grains in stycast epoxy \cite{klaassen}. The SiC
grains create a rough surface to prevent an angular dependence of
reflection, and the coating has an absorptivity of 90 \% in the
0.3-2.5 THz range \cite{klaassen}. The coax filters have a 50
$\Omega$ impedance, and use bronze and carbon powder as absorber
along with a NbTi central conductor, following Ref. \cite{milliken}.
At 4.2 K, the transmission up to 20 GHz is given by $S = Af$, with
$A=-0.18$ dB/GHz. At 4 GHz the attenuation is below 1 dB, while we
estimate that 4.2 K radiation is reduced by $\sim 30$ dB to a power
below 100 fW, which is an upper limit due to additional absorption
by the carbon. The sample stage is attached to the 50 mK cold finger
of the ADR. For readout, we use two 0.86 mm diameter CuNi coaxial
cables, connected between the 4 K stage and the coax filters on the
sample stage.

Having shown that the new design reduces stray light and improves
resonator quality factors, we next describe the influence of key
parts of the setup. We test the effect of the: I) outer can, II)
black coating on the outer can and/or sample box lid, III) coax
filters, and IV) seams in the outer box. The influence of stray
light is quantified by continuously measuring the quality factors of
the resonators while warming up the cryostat at the 4~K stage,
bathing the sample stage in a hot thermal photon bath. While doing
so, the sample stage temperature is always kept below 150 mK where
the quality factor is unaffected.

\begin{figure}[t]
    \centering
    \includegraphics[width=1\linewidth]{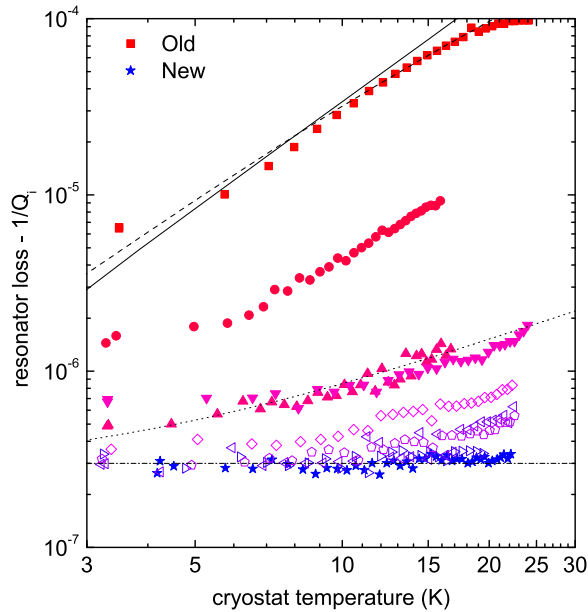}
    \caption{(Color online) The loss of an Al resonator versus cryostat temperature,
    showing the influence of different shielding techniques when going from no infrared shielding
    ($\blacksquare$)
    to fully shielded (\ding{72}).
    The sample stage temperature is kept below 150 mK.
    Variations in the presence of a coated sample box lid and coax filters (closed symbols, red colors):
    no can present (\ding{108}), sample box covered with Al tape
    ($\blacktriangle$)
    and uncoated can present
    ($\blacktriangledown$).
    Variations in the presence of a coated outer can (open symbols, purple and blue colors):
    coated can floating on 0.5 mm spacers ($\Diamond$),
    uncoated sample box lid and no coax filters (\pentagon), uncoated lid and filters ($\lhd$) coated lid and no filters ($\rhd$).
    Influence of a hot blackbody (solid), hot blackbody filtered with a cut-off frequency at 1 THz (dashed),
    filtered with a cut-off frequency at 100 GHz (dotted) and no dependence (dash dotted).}
    \label{figure:fig3}
\end{figure}

The effectiveness of the shielding methods is shown in Fig.
\ref{figure:fig3}. When not using any shielding (outer can, coax
filters, or a coated sample box lid), a loss of $10^{-5}$ is found,
which increases strongly with increasing cryostat temperature (red
squares). When adding a coated sample box lid and coax filters, a
loss on the order of $10^{-6}$ is found at the lowest cryostat
temperatures (red dots). However, here the loss also increases with
elevating cryostat temperature. Light-tightness is somewhat improved
when adding an uncoated can or covering the sample box with Al tape
(purple triangles). The largest improvement is observed when using a
coated can, although a small cryostat temperature dependence is
still visible. Only when using a coated can and a coated sample box
lid is the lowest loss achieved and the dependence on the
cryostat temperature fully removed (blue stars). When a 0.5 mm gap
is introduced a small temperature dependence returns. We find that
the effect of the coax filters on the light-tightness is
insignificant for our experiment.

The data follow only approximately a pure blackbody radiation
dependence for the unshielded case. For a pure blackbody: $P \propto
T^4$, hence $1/Q \propto T^2$ (solid line). With increasing
shielding the slope of the data decreases, consistent with the
sample stage acting as a low pass filter. We model this as a first
order filter with transfer function: $1/(1+[f/f_c]^2)$, with cut-off
frequency $f_c$. We find that for the unshielded case $f_c$ is on
the order of 1 THz (dashed line) (Eq. \ref{eq:lossvsnqp}). With Al
tape or an uncoated can: $f_c \sim 100$~GHz (dotted line). With each
additional shielding step the loss drops
--- indicating enhanced insensitivity to stray light --- and the
slope decreases --- indicating a decrease of the cut-off frequency.
This suggests that stray low frequency photons are the main source
of infrared-related loss in partly shielded environments.

The data in Fig. \ref{figure:fig3} demonstrate that a box-in-a-box
design with black coating is needed to ensure the removal of the
influence of stray light, and that anything less is insufficient. An
uncoated outer can increases the quality factor to above $10^6$ at
3~K, but does \emph{not} completely remove the influence of stray
light. Only when using an outer can and coating the inner surfaces
of the can and sample box is there no dependence on the cryostat
temperature. This temperature is varied from 3~K to 23~K, increasing
the stray light power by $10^3$. Moreover, a tight fitting of the
outer can is unnecessary as a 0.5 mm gap has only a small effect on
the quality factors at 3~K. Using a coated can is more effective
than having it tightly fitting. The coax filters are insignificant
for our mount, possibly due to the use of dissipative CuNi coaxial
lines. However, the filters are still useful as they ensure
thermalization of the inner wire.

\begin{figure}[t]
    \centering
    \includegraphics[width=1\linewidth]{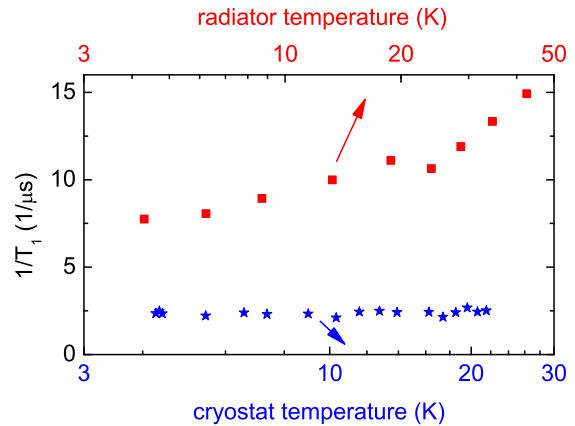}
    \caption{(Color online) Phase qubit energy relaxation rate measured in the light-tight setup versus cryostat temperature (\ding{72}),
    and measured without the presence of a coated outer can versus temperature of a radiator ($\blacksquare$) (see Fig. \ref{figure:fig2}).
    The sample stage temperature is kept below 150 mK.}
    \label{figure:fig4}
\end{figure}

The resonator quality factors improve from $10^5$ to $2 \cdot 10^6$,
as shown in Fig. \ref{figure:fig1} (closed symbols). This value is
believed to be unrelated to stray light because of its insensitivity
to the cryostat temperature. We estimate a lower limit of $10^8$ due
to stray light, assuming a level equal to the noise at 23~K in Fig.
\ref{figure:fig3} and extrapolating to 3~K. The remaining loss
mechanism may be radiation loss or excess quasiparticles from some
other mechanism, as suggested by recent number fluctuation
measurements \cite{visser2011}. In this case the quasiparticle
density has been reduced to 10 $\mu$m$^{-3}$ (blue solid line) or
below.

In order to quantify the influence of stray light on qubits, we also
measure the energy relaxation time $T_1$ of a phase qubit versus
mounting method. With the filters and black outer can in place, we
find a $T_1$ of 450 ns, consistent with typical values for phase
qubits. In addition, we find no increase of the energy relaxation
rate with increasing cryostat temperature, as shown in Fig.
\ref{figure:fig4}. In contrast, when only the outer can is removed
$T_1$ drops to 120 ns. This value is compatible with a quasiparticle
density of 170 $\mu$m$^{-3}$ \cite{rate}, close to the value of 230
$\mu$m$^{-3}$ found for the resonators. Without an infrared shield,
$T_1$ decreases very rapidly with increasing cryostat temperature.
To identify the influence of stray infrared light on the qubit, we
instead use a blackbody radiator, which is placed behind the
magnetic shield (see Fig. \ref{figure:fig2}) and heated up to a
stable temperature. We emphasize that the radiator has a weaker
influence than the cryostat. The energy relaxation rate clearly
increases with the radiator temperature. The decrease in $T_1$ as
well as the temperature dependence in Fig. \ref{figure:fig4} show
that stray light considerably diminishes qubit coherence. It is
therefore vitally important for qubit coherence to use a light-tight
sample stage, as shown in Fig. \ref{figure:fig2}.

In summary, when placing a superconducting resonator or qubit in a
simple mounting box, the quality factor and energy relaxation times
can be significantly influenced by stray infrared light. Moreover,
stray light is often the dominant limitation in present experiments,
introducing an excess quasiparticle density between 170-230
$\mu$m$^{-3}$. Using a combination of shielding methods we have
improved quality factors from $10^5$ to $2\cdot 10^6$ and qubit
energy relaxation times from 120 to 450~ns, with measurements now
being unaffected by stray light. This shows that the influence from
stray infrared can be removed using a `box-in-a-box' design with
black absorbers. We estimate a lower limit of $10^8$ for resonator
quality factors due to stray light in the present configuration.

\begin{acknowledgments}
This work was supported by IARPA under ARO award W911NF-09-1-0375
and by the Rubicon program of the Netherlands Organisation for
Scientific Research (NWO).
\end{acknowledgments}

\end{document}